\documentclass[preprint,showpacs,preprintnumbers,amsmath,amssymb]{revtex4}

\usepackage{graphicx}
\usepackage{epsfig}
\usepackage{bm}
\usepackage{amsfonts}
\usepackage{subfigure}
\usepackage{multirow}
\usepackage{color}
\usepackage{float}

\newcommand*{\D}{{\rm d}}

\newcommand*{\mpl}{M_{\rm pl}}

\begin{document}



\title{\textbf{Primordial black holes and induced gravitational waves in $k$-inflation}}



\author{Milad Solbi\footnote{miladsolbi@gmail.com} and Kayoomars Karami\footnote{kkarami@uok.ac.ir}}
\address{\small{Department of Physics, University of Kurdistan, Pasdaran Street, P.O. Box 66177-15175, Sanandaj, Iran}}

\date{\today}

\begin{abstract}
Recent observational constraints indicate that primordial black holes (PBHs) with the mass scale $\sim 10^{-12}M_{\odot}$ can explain most of dark matter in the Universe. To produce this kind of PBHs, we need an enhance in the primordial scalar curvature perturbations to the order of ${\mathcal{O}(10^{-2})}$ at the scale $ k \sim 10^{12}~\rm Mpc^{-1}$. Here, we investigate the production of PBHs and induced gravitational waves (GWs) in the framework of \textbf{$k$-inflation}. We solve numerically the Mukhanov-Sasaki equation to obtain the primordial scalar power spectrum. In addition, we estimate the PBHs abundance $f_{\text{PBH}}^{\text{peak}}$ as well as the energy density parameter $\Omega_{\rm GW,0}$ of induced GWs. Interestingly enough is that for a special set of model parameters, we estimate the mass scale and the abundance of PBHs as $\sim{\cal O}(10^{-13})M_{\odot}$ and $f_{\text{PBH}}^{\text{peak}}=0.96$, respectively. This confirms that the mechanism of PBHs production in our inflationary model can justify most of dark matter. Furthermore, we evaluate the GWs energy density parameter and conclude that it behaves like a power-law function $\Omega_{\rm GW}\sim (f/f_c)^n$ where in the infrared limit $f\ll f_{c}$, the power index reads $n=3-2/\ln(f_c/f)$.
 \end{abstract}



\maketitle

\newpage
\section{Introduction}
Primordial black holes (PBHs) have been a topic of debate for years \cite{Hawking:1971,Carr:1974,Carr:1975}. The existence of PBHs was proposed by Zel'dovich and Novikov for the first time in 1967 \cite{zeldovich:1967}. PBHs received new attentions after the LIGO-Virgo team detected gravitational waves (GWs) from a binary black hole (BH). After the first detection, the LIGO-Virgo collaboration observed some of these GW events from binary BHs  \cite{Abbott:2016-a,Abbott:2016-b,Abbott:2017-a,Abbott:2017-b,Abbott:2017-c}. They found that the mass of these BHs is nearly $ 30 M_\odot $. The stellar BHs do not fall on this mass scale. However, PBHs that have a wide mass range can explain this type of BHs. In other words, PBHs can be a candidate for the origin of BHs that was discovered by LIGO-Virgo collaboration \cite{Bird:2016,Clesse:2017,Sasaki:2016,Carr:2016}.

PBHs with a $12-213~ {\rm Gpc^{-3} yr^{-1}}$ merger rate and mass about $ 10 M_{\odot} $ have good consistency with the LIGO-Virgo observations \cite{Abbott:2017-a}. Sasaki et al. \cite{Sasaki:2016} showed that PBHs with this merger rate can constitute approximately ${\mathcal{O}(1)}\% $ of the total dark matter (DM). Recently in the  data, six microlensing events were detected, which have ultra-short timescales \cite{OGLE2}. The results of this investigation show that PBHs with a mass scale of ${\mathcal{O}(10^{-5})}M_\odot $ and the abundance around ${\mathcal{O}(10^{-2})}$ can locate in the allowed region of OGLE data \cite{OGLE2}.

Following the recent observations of white dwarf explosion (WD) \cite{WD} and microlensing events with Subaru HSC (Subaru HSC) \cite{HSC,HSC-b}, it is possible that the PBHs with the mass range ${\mathcal{O}(10^{-13})}M_\odot \leq M \leq {\mathcal{O}(10^{-11})}M_\odot$ can describe all DM in the universe.

The collapse of primordial curvature perturbations after horizon reentry can form PBHs. For PBHs production, the amplitude of primordial curvature perturbations needs to be increased by nearly seven orders to $A_{s}\sim {\mathcal{O}(10^{-2})} $ \cite{sato:2019}. At the pivot scale $k_{*}=0.05{\rm Mpc}^{-1} $, the exact value of the primordial curvature perturbations has been estimated by cosmic microwave background (CMB) observations. The CMB anisotropy measurements show $A_{s}=2.1 \times 10^{-9}$ \cite{akrami:2018}. The enhancement of the curvature perturbations leads to an overdense region. After horizon reentry, these overdense areas can generate considerable metric perturbations in addition to PBHs. In contrary to the linear level, scalar and tensor perturbation couplings occur at the second order. The scalar metric perturbations can produce the stochastic GW background through the second-order effect \cite{Kohri:2018,Cai:2019-a,HSC-b,Bartolo:2019-a,Bartolo:2019-b,Wang:2019,Cai:2019-c,Xu:2020,Lu:2019,Hajkarim:2019,Domenech:2020a,Domenech:2020b,Fumagalli:2020b}. Hence, induced GWs signal detection suggests a new way of looking for the existence of PBHs.

More recently, in the context of inflation, different mechanisms of PBHs production have been proposed, see e.g. \cite{Cai:2018,Ballesteros:2019,Ballesteros:2020a,Ballesteros:2020b,Kamenshchik:2019,Inomata:2018,Motohashi:2017,Ezquiaga:2018,Germani:2017,Di:2018,Ballesteros:2018,Dalianis:2019,Dalianis:2020,
chen:2019,Ozsoy:2018,Tada:2019,Liu:2020,Bellido:2017,mishra:2020,Fumagalli:2020a,Atal:2019,Khlopov:2010,Belotsky1:2014,Belotsky:2019,shiPi:2018,Braglia:2020,Braglia2:2020,mahbub:2020}.
For instance, in the single field model of inflation with an inflection point, the primordial curvature power
spectrum can be enhanced at small scales \cite{Motohashi:2017,Germani:2017,Di:2018}. Indeed, in the vicinity of inflection point, the inflaton experiences the ultra slow-roll (USR) phase. Then, the inflaton velocity decreases in the USR phase due to high friction and curvature perturbation increases consequently.

In the context of inflation, the Galileon scalar field scenario is one of the most attractive models \cite{Tsujikawa:2011,Tsujikawa:2011-b,teimoori:2018,kobayashi:2010,Ohashi:2012,Tumurtushaa:2019,Burrage:2010}.
The Galileon inflation takes place in a sub-class of the Horndeski theory \cite{Horndeski:1974}. Hirano et al. \cite{Hirano:2016} pointed out that Galileon inflation models can explain CMB anomalies when the inflaton goes into the USR phase. Also, they showed that the Galileon inflation in the USR phase can illustrate the decaying of CMB power spectrum at the largest scales. Moreover, the introduction of an appropriate Galileon term could enhance the scalar power spectrum and lead to PBHs. Furthermore, another interesting inflationary model is known as the $k$-inflation \cite{Armendariz:1999,Garriga:1999}. Such inflationary model is driven by non-minimal kinetic terms, which can originate from low-energy effective string theory action \cite{Garriga:1999}.

In this paper, we investigate the possibility of PBH formation in the $k$-inflation  scenario when the base inflationary potential has an exponential form. The outline of the paper is as follows. We review $G$ and $k$ inflationary models in Sec. \ref{sec2}. The mechanism of PBH formation is explained in Sec. \ref{sec3}. In Sec. \ref{sec5}, we calculate the abundance of PBHs. The induced GWs are studied in Sec. \ref{sec6}. Also Sec. \ref{sec7} is devoted to our conclusions.
\section{$G$ \MakeLowercase{and} $k$ Inflation}\label{sec2}
The general action of \textbf{$G$-inflation} in Galileon scenario is given by  \cite{kobayashi:2010}
\begin{equation}\label{action}
S= \int {\rm d}^{4}x \sqrt{-g}\left[\frac{M_{\rm pl}^{2}}{2} R + {\cal L}_{\phi} \right],
\end{equation}
where
\begin{equation}\label{Lagrangian}
{\cal L}_{\phi}\equiv K(\phi, X)-G(\phi, X)\Box \phi,
\end{equation}
is the Lagrangian of the scalar field, and $g$ and $R$ are the determinant of the metric $g_{\mu \nu}$ and Ricci scalar, respectively. Also $K$ and $G$ are the general functions of the scalar field $\phi$ and the kinetic term $X\equiv -\frac{1}{2}g^{\mu\nu}\phi_{,\mu}\phi_{,\nu}$.

In the framework of Galileon gravity (\ref{action}), the Friedmann equations take the forms \cite{kobayashi:2010,Ohashi:2012}
\begin{equation}\label{Friedman1}
3 M_{\rm pl}^{2} H^{2}=2K_{,X}X-K+3G_{,X}H\dot\phi^3-2G_{,\phi} X,
\end{equation}
\begin{equation}\label{Friedman2}
-M_{\rm pl}^{2}\left(3H^2+2\dot H\right)=K-2\left(G_{,\phi}+G_{,X}\ddot\phi \right)X,
\end{equation}
where $,_{X}\equiv {\rm d}/{\rm d}X$ and $,_{\phi}\equiv {\rm d}/{\rm d}\phi$. Also the dot denotes the derivative with respect to the cosmic time. The scalar field equation of motion from action (\ref{action}) can be derived as
\begin{align}
 K_{,X} &\left(  \ddot{\phi}+3H\dot{\phi} \right) + 2 K_{,XX} X \ddot{\phi} +
   2K_{,X\phi} X - K_{,\phi}-2 \Big( G_{,\phi}-G_{,X\phi}X \Big) \left( \ddot{\phi}+3H\dot{\phi} \right)\nonumber \\
& + 6G_{,X} \Big[ \left( HX \right)\dot{}+3H^2 X \Big]- 4G_{,X\phi} X \ddot{\phi} - 2G_{,\phi\phi}X+6HG_{,XX}X \dot{X}=0.
\label{eqn:EOM}
\end{align}
The second order action for curvature perturbation ${\cal R}$ at the first order approximation is found to be \cite{kobayashi:2010}
\begin{eqnarray}\label{2ndaction}
S^{(2)}=\frac{1}{2} \int\D\tau\D^3x
z^2\left[{\cal G}({\cal R}_\phi')^2-{\cal F}(\Vec{\nabla}{\cal R}_\phi)^2\right],
\end{eqnarray}
in which
\begin{eqnarray}\label{FG-form}
z&\equiv&\frac{a\dot\phi}{H-G_{,X}\dot\phi^3/2\mpl^2},
\\
{\cal G}&\equiv&K_{,X}+2XK_{,XX}+6G_{,X}H\dot\phi+6\frac{G_{,X}^2}{\mpl^2}X^2-2\Big( G_{,\phi}+XG_{,\phi X} \Big)+6G_{,XX}HX\dot\phi,\\
{\cal F}&\equiv&K_{,X}+2G_{,X}\left(\ddot\phi+2H\dot\phi \right)
-2\frac{G_{,X}^2}{\mpl^2}X^2+2G_{,XX}X\ddot\phi -2 \Big(G_{,\phi}-XG_{,\phi X}\Big),
\end{eqnarray}
and the prime shows the derivative with respect to the conformal time $\tau$. According to \cite{kobayashi:2010}, the squared sound speed is obtained as $c_s^2={\cal F}/{\cal G}$, where the conditions ${\cal F} > 0$ and ${\cal G }> 0$ are needed to avoid of ghost and gradient instability.
\\
The power spectrum of ${\cal R}$ produced during the $G$-inflation can be assessed in the Fourier space by writing the Mukhanov-Sasaki (MS) equation as
\begin{eqnarray}\label{MS_Eq}
\frac{\D^2u_k}{\D y^2}+\left(k^2-\frac{\tilde z_{,yy}}{\tilde z}\right)u_k=0,
\end{eqnarray}
where
$\D y=c_s\,\D \tau$, $\tilde z\equiv\left({\cal F}{\cal G}\right)^{1/4}z$, and
$u_k\equiv\tilde z{\cal R}_{\phi,k}$ \cite{kobayashi:2010}. Also the scalar power spectrum is given by
\begin{equation}
{\cal P}_{\cal R}(k)=(2\pi^{2})^{-1}k^{3}\vert u_{k}/z\vert^{2}.
\end{equation}
Hereafter, in the Lagrangian (\ref{Lagrangian}) we take
\begin{equation}
K(\phi,X)=X-V(\phi).
\end{equation}
Also following \cite{lin:2020}, we assume that $G$ depends only on $\phi$, and consequently, we can convert the term $G(\phi)\Box \phi$ into $-2g(\phi)X$ via partial integration in action (\ref{action}), where $g(\phi)=d G(\phi)/d\phi$. Hence, the  action (\ref{action}) is turned to
\begin{equation}
\label{action2}
S=\int {\rm d}^4x\sqrt{-g}\left[\frac{M_{\rm pl}^{2}}{2}R+\big(1-2g(\phi) \big)X-V(\phi)\right].
\end{equation}
One can consider the action (\ref{action2}) as a limited class of $k$-inflation models, where the kinetic term depends on the scalar field only \cite{Barenboim:2007,lin:2020}.
Hence, Eqs. (\ref{Friedman1})-(\ref{eqn:EOM}) are reduced to
\begin{gather}
\label{eom1}
3H^2=\frac{1}{2}\dot{\phi}^2+V(\phi)-\dot{\phi}^2g(\phi),\\
\label{eom2}
2\dot{H}+3H^2+\frac{1}{2}\dot{\phi}^2-V(\phi)-\dot{\phi}^2g(\phi)=0,\\
\label{eom3}
\ddot{\phi}+3H\dot{\phi}+\frac{V_{\phi}-\dot{\phi}^2g_{,\phi}}{1-2g(\phi)}=0.
\end{gather}

Now, we are interested in investigating the \textbf{$k$-inflation} in the slow-roll regime. In the slow-roll approximation $\varepsilon \equiv - \dot{H}/H^2\ll 1$ and $\eta \equiv - \ddot{\phi}/(H \dot{\phi}) \ll 1$, one can easily show that the background Eqs. (\ref{eom1}) and (\ref{eom3}) are simplified into \cite{lin:2020}
\begin{gather}
  3H^2\simeq V, \\
  3H\dot{\phi}(1-2g)+V_{\phi}\simeq0.
\end{gather}
In the slow-roll regime, the power spectrum of curvature perturbations and  primordial gravitational waves are given by \cite{lin:2020}
\begin{equation}
\label{ps-sr1}
{\cal P}_{\cal R}=\frac{H^2}{8\pi^2\varepsilon}\simeq\frac{V^3}{12\pi^2V_{\phi}^2}(1-2g),\\
\end{equation}
and
\begin{equation}
\label{ps-sr2}
{\cal P}_T=\frac{H^2}{2\pi^2},
\end{equation}
respectively. Consequently, the scalar spectral index and the tensor-to-scalar ratio can be obtained as
\begin{equation}
\label{nsG3}
n_s-1=\frac{1}{1-2g}\left(2\eta_V-6\varepsilon_V+
\frac{2g_{\phi}}{1-2g}\sqrt{2\varepsilon_V}\right),
\end{equation}
\begin{equation}
\label{rG3}
r=\frac{{\cal P}_T}{{\cal P}_{\cal R}}=\frac{16 X(1-2g)}{H^2},
\end{equation}
where $\varepsilon_{\rm v}\equiv\frac{M_{\rm pl}^2}{2}\left( \frac{V_{,\phi}}{V} \right)^2 $ and $\eta_{\rm v}\equiv  M_{\rm pl}^2 \left(\frac{V_{,\phi \phi}}{V}\right)$ are defined as potential slow-roll parameters. In the next section, we present the results of exact numerical solutions obtained for the  background equations (\ref{eom1}), (\ref{eom3}) as well as the MS equation (\ref{MS_Eq}). We use the slow-roll solutions as initial conditions for our numerical calculations.
\section{PBH Formation Mechanism}\label{sec3}
Enhancing the scalar power spectrum at small scales can achieve by introducing an adequate kinetic term.
The intended function must cause the model to be compatible with the Planck observations at the CMB scale and explain the PBHs formation at small scales simultaneously.
Hence, the challenge is finding a proper kinetic term.
A proposed solution for managing the challenge is dividing the $g(\phi)$ into two sections as follows \cite{Dalianis:2020}
\begin{equation}\label{g}
g(\phi)=g_I(\phi)\Big(1+g_{II}(\phi)\Big),
\end{equation}
where
\begin{equation}\label{gI}
g_I(\phi)=-e^{\alpha \, \phi/M_{\rm pl}},
\end{equation}
\begin{equation}\label{gII}
g_{II}(\phi)=\frac{d}{\sqrt{\left(\frac{\phi-\phi_c}{c}\right)^2+1}}\,.
\end{equation}
The compatibility of the model with the Planck measurements is obtained by $g_I(\phi)$. In other words, $g_I(\phi)$ is responsible for producing quantum fluctuations consistent with the CMB constraints on $n_{s}$ and $r$. On the other hand, $g_{II}(\phi)$ may yield the scalar curvature successfully increases to generate PBHs at $\phi=\phi_c $. Also, at distances far from  $\phi=\phi_c $, the value of $g_{II}(\phi)$ goes down. Here, it is unavoidable to notice that $\alpha$ and $d$ are dimensionless variables, while $\phi_c $ and $c$ have the mass dimension. Additionally, $d$ and $c$ represent the height and width of the peak of the function $g_{II}(\phi)$, respectively.

In the present work, we consider the exponential potential as the base potential which is given by
\begin{equation}\label{vtotal}
V(\phi)=V_{0}\,e^{\lambda \, \phi/M_{\rm pl}},
\end{equation}
where $V_{0}$ is fixed by $P_{\cal R}(k_{*})=2.1\times 10^{-9}$ and $k_{*}=0.05~{\rm Mpc}^{-1}$ denotes the pivot scale. In standard canonical inflation, the exponential potential does not satisfy the Planck observation at the CMB scale \cite{karami:2017}. Also, inflation never ends in the standard model, when the potential has exponential form \cite{karami:2017}. Both of these problems inspire us to investigate the potential (\ref{vtotal}) in the $k$-inflation scenario.

Therefore, our model is characterized by a set of six parameters (i.e. $\alpha$, $d$, $c$, $\phi_{c}$, $V_{0}$, and $\lambda$). We set $N_{*}=0$ at the horizon exit and take the slow-roll value for the scalar filed as $\phi_{*}\simeq 0.173 $ at the pivot scale $k_{*}=0.05{\rm Mpc}^{-1} $. We further consider $\alpha=50$ and $\lambda=8$, which lead to $n_{s}\in (0.956,0.978)$ and $r\leq 0.06$ and can satisfy the CMB bounds on $n_{s}$ and $r$.
Moreover, the parameter $V_{0}$ is fixed by the power spectrum at the CMB scale. Therefore, $d$, $c$, and $\phi_c$ are the only remaining parameters in our model which can affect the both scalar perturbation evolution and PBH production.

It is necessary to notice that,
at first glance, the action (\ref{action2}) could seem to be a minor improvement of the canonical model, because in such a circumstance, the non-canonical kinetic term can be converted into a canonical term, using a field redefinition like
\begin{eqnarray}\label{transform}
 \psi=\int{\sqrt{1-2 \it{g} (\phi)}}~ {\rm d}\phi.
\end{eqnarray}
Therefore, it can be possible to express the non-canonical term as a canonical scalar field $\psi$ with a potential $U(\psi)=V(\phi(\psi))$.
Nevertheless, some models that appear to be complex in the canonical framework may have a simple interpretation in the non-canonical equivalent representation. Hence, it is plausible  to investigate them on a non-canonical basis \cite{Barenboim:2007}.

Furthermore, recent parers \cite{lin:2020,zhang:2021,yi:2021} investigated the action (\ref{action2}) in the non-canonical framework. The non-canonical term in their model is complicated, so that using the transform equation may lead to more complexity. Hence, they perform the calculations in the non-canonical framework for simplicity. Also, in \cite{yi:2021}, the authors utilize Eq. (\ref{transform}) at large scales to find the potential form in the pivot scale. As a result, the values of $n_{\rm s}$ and $r$  can be calculated most simply.

In our model, the peak function $g_{II}(\phi)$ is complicated. If the transformation contains it, the potential in the canonical action will be very complex. On the other hand, the value of the $g_{II}(\phi)$ is negligible at far away from the peak, and as a result one can consider the $g_I(\phi)$ to change the non-canonical term to the canonical form at the large scales. In addition,  $g_I(\phi) \gg 1$ so that the transformation equation is reduced to $\psi=\int{\sqrt{-2 \it{g_I} (\phi)}}~ {\rm d}\phi$ in this special case. Now, we can redefine scalar field as $\psi=\frac{1}{\alpha}\ln{(\frac{\alpha \phi}{ \sqrt{2}})}$ to the action (\ref{action2}) turns to the canonical form. Also, for the new scalar filed $\psi$, the potential $V(\phi)$ turns to $U(\psi)=U_{0} \psi^{\lambda/\alpha}$, where $U_{0}\equiv 2^{-\frac{\lambda	}{2\alpha} } \alpha^{\frac{\lambda}{\alpha} }V_{0} $ is a constant. In addition, $g_I(\phi)$ is dominated at the end of inflation so that $U(\psi)$ can consider as the potential of this era. According to the values of $\lambda=8$ and $\alpha=50$, the potential $U(\psi)$  does not have a minimum value, and hence reheating mechanism would not proceed in a usual way.

As mentioned above, our model would be more complicated when we consider the complete form of the $g(\phi)$ in transforming equation (\ref{transform}). Hence, we performed the calculations in the non-canonical framework to avoid of complexity.

\begin{figure}[h]
\begin{minipage}[b]{1\textwidth}
\subfigure[\label{fig-phi} ]{ \includegraphics[width=.46\textwidth]%
{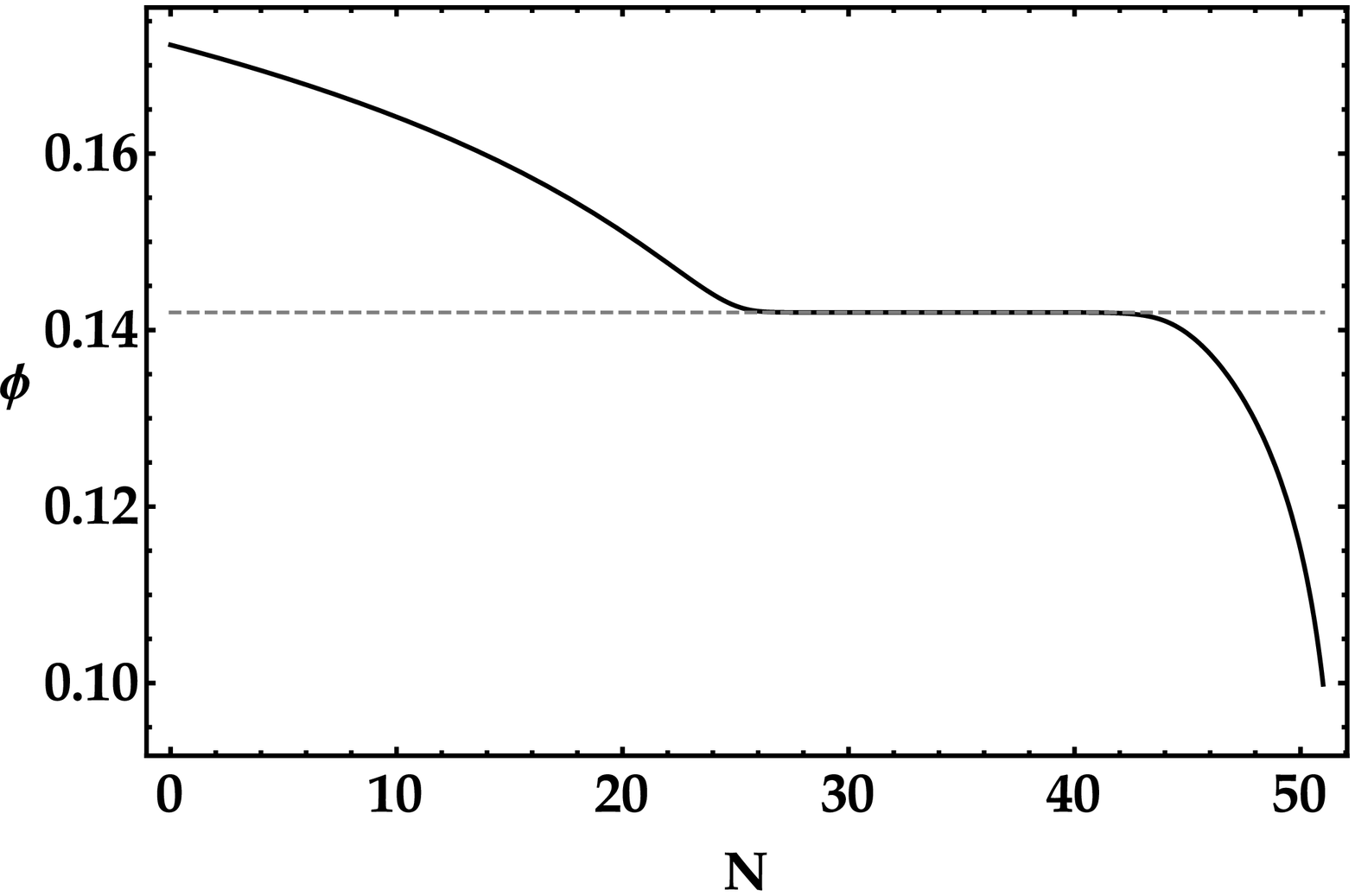}}\hspace{.1cm}
\subfigure[\label{fig-srp}]{ \includegraphics[width=.46\textwidth]%
{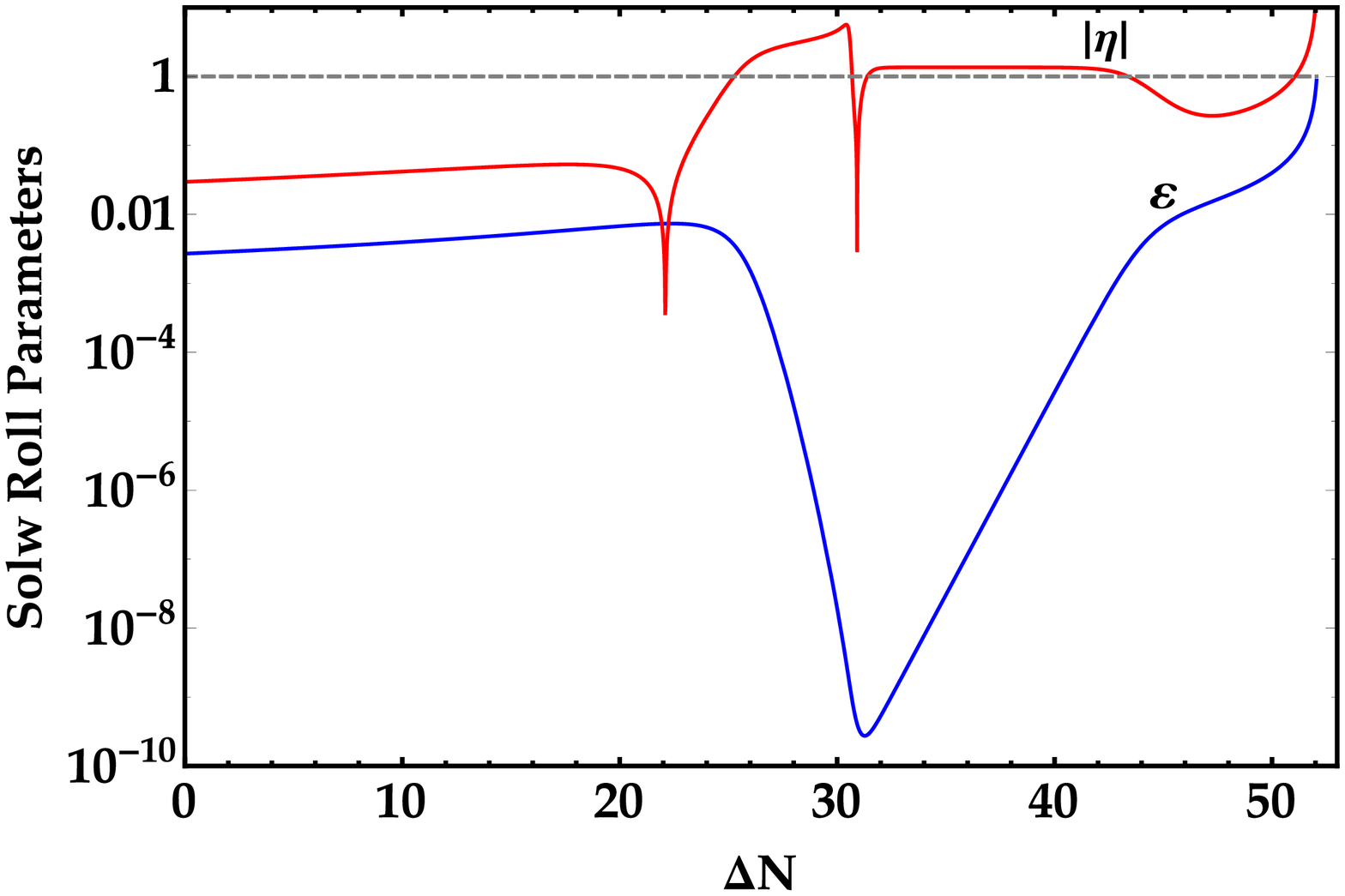}}
\end{minipage}
\caption{(a) Evolution of the scalar filed $\phi$ versus the $e$-fold number $N$. The dashed line represents $\phi=\phi_c$. (b) Evolution of the slow-roll parameters $\varepsilon$ and $\eta$ in terms of $N$. The dashed line shows the violation of the slow-roll condition. The auxiliary parameters are given by the parameter set $C$ in Table \ref{tab1} and we take $N_{*}=0$ at the horizon exit.
  }\label{linear}
\end{figure}

Note that beyond the slow-roll conditions, Eq. (\ref{ps-sr1}) does not hold and we need to solve the MS equation (\ref{MS_Eq}) numerically to obtain the scalar power spectrum ${\cal P}_{\cal R}(k)=(2\pi^{2})^{-1}k^{3}\vert u_{k}/z\vert^{2}$. To do this, for the three sets of parameters listed in Table \ref{tab1} we could find the corresponding power spectrum with suitable enhance in PBH production as shown in Table \ref{tab2}. In Fig. \ref{fig-phi}, evolution of the scalar field versus the $e$-fold number is presented for the parameter set $C$. The flat part in Fig. \ref{fig-phi} is due to the increasing friction in this region, where the slow-roll condition is violated and inflaton is passing the USR phase. Hence, the scalar curvature perturbations intensively enhance. As shown in Fig. \ref{fig-srp}, the value of $\eta$ becomes larger than 1, and the slow-roll condition is not respected.
The sharp decrease in $\varepsilon$ can provide considerably increase in the scalar power spectrum.
\begin{table}[H]
  \centering
  \caption{The selected parameter sets for PBHs production}
  \begin{tabular}{cccc}
  \hline
  Sets \quad &\quad $\phi_{c}$ \quad & \quad $d$\quad &$c$\quad \\ [0.5ex]
  \hline
  \hline
$A$ \quad &\quad $0.167$ \quad &\quad $5.64\times 10^{7}$ \quad &\quad $10^{-11}$ \quad\\[0.5ex]
  \hline
$B$ \quad &\quad $0.159$ \quad &\quad $8.24\times 10^{7}$ \quad &\quad $10^{-11}$ \quad\\[0.5ex]
  \hline
$C$ \quad &\quad $0.142$ \quad &\quad $1.98\times 10^{8}$ \quad &\quad $10^{-11}$ \quad\\[0.5ex]
  \hline
  \end{tabular}
  \label{tab1}
\end{table}

\begin{table}[H]
  \centering
  \caption{The values of ${\cal P}_{\cal R}^\text{peak}$, $f_{\text{PBH}}^{\text{peak}}$ and $M_{\text{PBH}}^{\text{peak}}$ for three cases of Table \ref{tab1}.}
  \begin{tabular}{ccccccc}
  \hline
  Sets \quad & \quad $n_{s}$\quad &$r$\quad & \quad$k_{\text{peak}}/\text{\rm Mpc}^{-1}$ \quad &\quad $M_{\text{PBH}}^{\text{peak}}/M_{\odot}$ \quad & \quad ${\cal P}_{\cal R}^\text{peak}$\quad & \quad $f_{\text{PBH}}^{\text{peak}}$\\ [0.5ex]
  \hline
  \hline
$A$ \quad &\quad $0.972$ \quad &\quad $0.041$ \quad &\quad $2.63\times10^{5}$ \quad &\quad $36.19$ \quad &\quad $0.059$ \quad &$0.002$\\[0.5ex]
  \hline
$B$ \quad &\quad $0.964$ \quad &\quad $0.042$ \quad &\quad $3.77\times10^{8}$ \quad &\quad $1.75\times10^{-5}$ \quad &\quad $0.043$ \quad & $0.022$ \\[0.5ex]
  \hline
$C$ \qquad &\quad $0.961$ \quad &\quad $0.043$ \quad &\quad $2.86\times10^{12}$ \quad &\quad $3.06\times 10^{-13}$ \quad &\quad $0.033$ \quad &$0.96$ \\[0.5ex]
  \hline
  \end{tabular}
  \label{tab2}
\end{table}


\begin{figure}[H]
\centering
\includegraphics[scale=0.5]{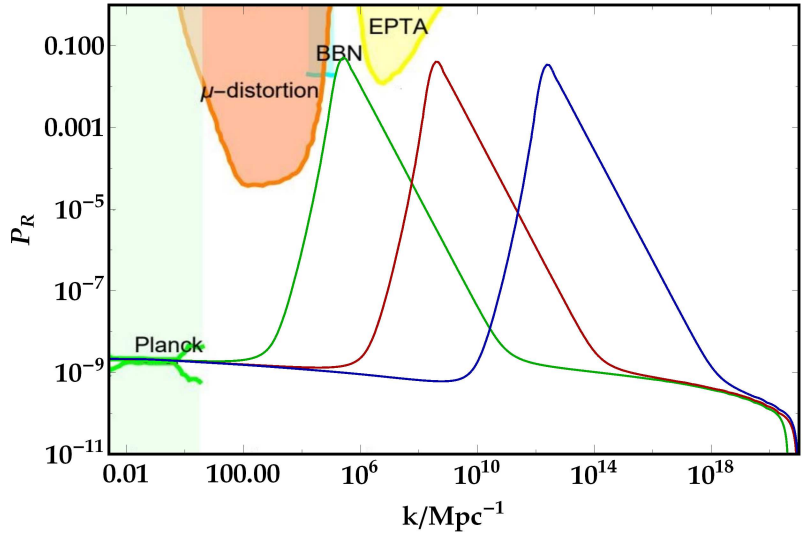}
\vspace{-0.5em}
\caption{The scalar power spectrum ${\cal P}_{\cal R}$ in terms of wavenumber $k$. The green, red and blue lines are corresponding to the parameter sets $A$, $B$ and $C$, respectively. The CMB observations exclude the light-green shaded area \cite{akrami:2018}. The yellow region demonstrates the constraint from the PTA observations \cite{Inomata:2019-a}. The cyan area represents the effect on the ratio between neutron and proton during the big bang nucleosynthesis (BBN) \cite{Inomata:2016}. The orange zone shows the $\mu$-distortion of CMB \cite{Fixsen:1996}.}
\label{fig-pr}
\end{figure}

\vspace{-1.2cm}
~~\\
In Fig. \ref{fig-pr}, the scalar power spectrum is plotted for the all parameter sets listed in Table \ref{tab1}. The results show that at the CMB scale (i.e. $k\sim0.05~ \rm Mpc^{-1}$),  the power spectrum is in order of ${\cal O}(10^{-9})$ which is consistent with the observations. Figure presents that the PBHs are generated when the peak of the scalar power spectrum increases by nearly seven orders of magnitudes, i.e. ${\cal P}_{\cal R}^\text{peak}\sim{\cal O}(10^{-2})$, at small scales. Our result for the PBHs production is in well agreement with the constraints deduced from the observations of CMB $\mu$-distortion, big bang nucleosynthesis (BBN), and pulsar timing array (PTA) \cite{Inomata:2019-a,Inomata:2016,Fixsen:1996}.
\section{Abundance of primordial black holes}\label{sec5}
Here, we are interested in obtaining the abundance of PBHs. We consider that the primordial curvature perturbations re-enter the horizon during the radiation-dominated (RD) era. If primordial curvature perturbations undergo a significant rise, the gravity of this overdense region can overcome the RD pressure at the horizon re-entry. Consequently, the collapse of curvature perturbations can produce PBHs.

\begin{figure}[H]
\centering
\includegraphics[scale=0.55]{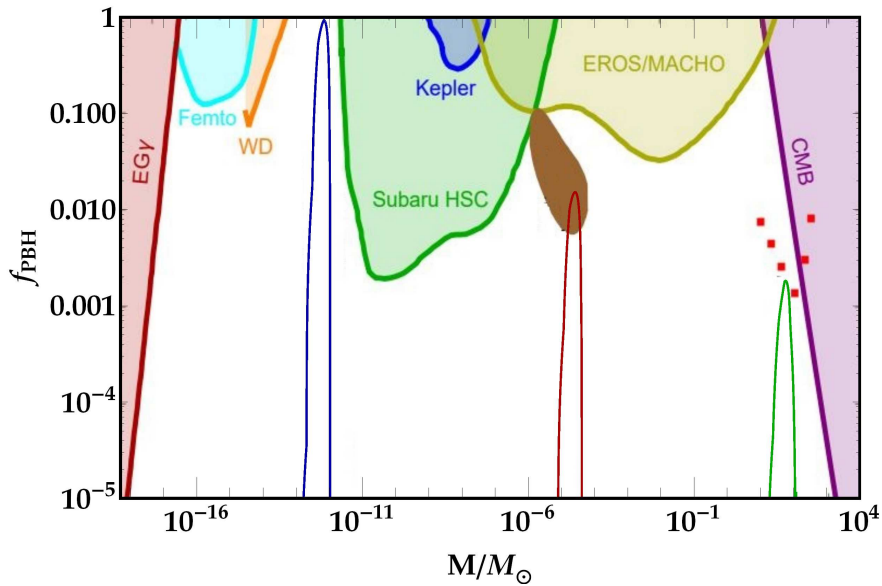}
\vspace{-0.5em}
\caption{The PBHs abundance $f_{\rm PBH}$ for the parameter sets $A$ (green line), $B$ (red line), and $C$ (blue line). The observational constraints on PBH abundance are shown by the shaded areas. The red dots show the upper bound on the PBH abundance due to the upper limit on the LIGO event merger rate \cite{Ali:2017}. The brown shaded area shows the allowed region of PBH abundance in the OGLE data \cite{OGLE2}. Other shaded regions illustrate the recent observational constraints including extragalactic gamma rays from PBH evaporation (EG$\gamma$) \cite{EGG}, femtolensing of gamma-ray burst (Femto) \cite{femto}, white dwarf explosion (WD) \cite{WD}, microlensing events with Subaru HSC (Subaru HSC) \cite{HSC,HSC-b}, with the Kepler satellite (Kepler) \cite{kepler}, with EROS/MACHO (EROS/MACHO) \cite{EROS}, and accretion constraints from CMB \cite{CMB-a,CMB-b}.}
\label{fig-fpbh}
\end{figure}

The PBH mass at the production time is given by $M=\gamma M_{\rm H}$, where $M_{\rm H}$ is the horizon mass \cite{Fu:2019,Sasaki:2018,Salopek:1998,Mukhanov:1991,Starobinsky:1992,Bellido:1996,Randall:1996,Ivanov:1994,Ivanov:1998}. Also the parameter $\gamma$ indicates the collapse efficiency, which relies on the characteristics of gravitational collapse. In this paper, we consider $\gamma=0.2$ \cite{Carr:1975,Sasaki:2018,Inomata:2017}. The ratio of the PBH mass to the total DM at the present time is given by  \cite{Sasaki:2018}
\begin{equation}\label{fpbheq}
f_{\rm PBH}(M_{\rm PBH})=1.68\times 10^{8} \left(\frac{\gamma}{0.2} \right)^{\frac{1}{2}} \left(\frac{g_{*}}{106.75} \right)^{-\frac{1}{4}} \left(\frac{M_{\rm PBH}}{M_{\odot}} \right)^{-\frac{1}{2} }\beta(M_{\rm PBH}),
\end{equation}
where $g_{*}$ indicates the effective degrees of freedom in the energy density during the PBH production process. Following \cite{Motohashi:2017}, we take $g_{*}=106.75$ to form the PBHs in the RD era. Also, $\beta$ is the mass fraction of PBH and can be calculated by  \cite{Sasaki:2018,young:2014,harada:2013}
\begin{equation}\label{betaeq}
\beta(M_{\rm PBH})=\gamma \frac{\sigma_{M_{\rm PBH}}}{\sqrt{2\pi}\delta_{\rm th}}\exp{\left(-\frac{\delta_{\rm th}^{2}}{2 \sigma_{M_{\rm PBH}}^{2}} \right)},
\end{equation}
where $\delta_{\rm th}$ is the threshold density contrast for PBH production. Many studies indicate that the numerically-approved threshold value may be very wide, with $\delta_{\rm th}$ ranging from $0.3$ to $0.66$ for PBH formation in the RD era \cite{harada:2013,Musco:2013,Germani:2019,Shibata:1999,Polnarev:2007,Musco:2009}. In the present work, we set $\delta_{\rm th} = 0.4$, which is consistent with the calculations in Refs. \cite{harada:2013,Musco:2013}.\\
Furthermore, $\sigma_{M_{\rm PBH}}$ in Eq. (\ref{betaeq}) stands for the variance of density contrast at the comoving horizon scale. The value of  $\sigma_{M_{\rm PBH}}$ is dependent on the mass of PBH $M(k)$ and is obtained as \cite{young:2014}
\begin{equation}\label{sigmaeq}
\sigma_{k}^{2}=\left(\frac{4}{9} \right)^{2} \int \frac{{\rm d}q}{q} W^{2}(q/k)(q/k)^{4} P_{\cal R}(q),
\end{equation}
where
$W(x)=\exp{\left(-x^{2}/2 \right)} $
is the Gaussian windows function. The mass of PBHs and the associated wavenumber can be related as \cite{Motohashi:2017,mishra:2020,Sasaki:2018}
\begin{equation}\label{masseq}
M_{\rm PBH}=1.13\times 10^{15}\left(\frac{\gamma}{0.2} \right) \left(\frac{g_{*}}{106.75} \right)^{-\frac{1}{6}}\left(\frac{k_{\rm PBH}}{k_{*}} \right)^{-2} M_{\odot}.
\end{equation}
The abundance of PBHs for each set of parameters in Table \ref{tab1} can be calculated using Eqs. (\ref{fpbheq}) and (\ref{masseq}). The results are displayed in Fig. \ref{fig-fpbh}. The shaded areas indicate the observational constraints on the PBH abundance. Figure \ref{fig-fpbh} shows that: (i) For the parameter set A, the peak of the PBH abundance is located at $36.19M_{\odot}$ and its value is approximately $f_{\text{PBH}}^{\text{peak}}\sim 0.002$. The achieved result for this case can describe the LIGO events, and it is compatible with the upper limit bounds on the LIGO merger rate. (ii) For the case B, the abundance peak occurs at $M_{\text{PBH}}^{\text{peak}}=1.75\times10^{-5}M_\odot$ and the mass spectrum height rises to $0.022$ which is located at the allowed region of microlensing events in the OGLE data. (iii) Interestingly enough is that for the set C, we obtain the mass scale for the PBHs as $\sim{\cal O}(10^{-12})M_{\odot}$ which lies in the observational region  ${\mathcal{O}(10^{-13})}M_\odot \leq M \leq {\mathcal{O}(10^{-11})}M_\odot$. Besides, we estimate the PBHs abundance as $f_{\text{PBH}}^{\text{peak}}=0.96$ which indicates the mechanism of the PBHs production can justify the most of DM in the Universe.

\section{Induced Gravitational Waves}\label{sec6}
The induced GWs can form simultaneously with PBHs when the large density disturbances re-enter the horizon in the RD era \cite{Kohri:2018,Cai:2019-a,Bartolo:2019-a,Bartolo:2019-b,Wang:2019,Cai:2019-c,Xu:2020,Lu:2019,Hajkarim:2019,Domenech:2020a,Domenech:2020b,HSC-b}. The  induced GWs can be tested by the future observations of PTA \cite{EPTA-a,EPTA-b,EPTA-c,EPTA-d} and space-based GWs observatory, such as LISA \cite{lisa}. In the conformal Newtonian gauge, the perturbed Friedmann-Robertson-Walker (FRW) metric has the following form \cite{Ananda:2007}
\begin{eqnarray}
ds^2=a(\eta)^2\left\{ -(1+2\Psi)d\eta^2 +\left[(1-2\Psi)\delta_{ij}+\frac{h_{ij}}{2} \right]dx^idx^j \right\}\;,
\end{eqnarray}
where $a$ and $\eta$ represent the scale factor and conformal time, respectively. Also $\Psi$ is the first-order scalar perturbations, and
$h_{ij}$  denotes the perturbation of the second-order transverse-traceless tensor. The inflaton would decay into light particles after the inflation to thermalize our universe once the reheating is over. Hence, the effect of the inflaton field is ignorable. Consequently, the standard Einstein equation will work properly during RD. Therefore, the second-order tensor perturbations $h_{ij}$ satisfy \cite{Ananda:2007,Baumann:2007}
\begin{eqnarray}\label{EOM_GW}
h_{ij}^{\prime\prime}+2\mathcal{H}h_{ij}^\prime - \nabla^2 h_{ij}=-4\mathcal{T}^{lm}_{ij}S_{lm}\;,
\end{eqnarray}
where  $\mathcal{H}\equiv a^{\prime}/a$ is the conformal Hubble parameter, and the quantity $\mathcal{T}^{lm}_{ij}$ is described as the transverse-traceless projection operator. The GW source term $S_{ij}$ is given by
\begin{eqnarray}
S_{ij}=4\Psi\partial_i\partial_j\Psi+2\partial_i\Psi\partial_j\Psi-\frac{1}{\mathcal{H}^2}\partial_i(\mathcal{H}\Psi+\Psi^\prime)\partial_j(\mathcal{H}\Psi+\Psi^\prime)\; .
\end{eqnarray}
The scalar metric perturbation $\Psi$ in the RD takes the form \cite{Baumann:2007}
\begin{eqnarray}
\Psi_k(\eta)=\psi_k\frac{9}{(k\eta)^2}\left(\frac{\sin(k\eta/\sqrt{3})}{k\eta/\sqrt{3}}-\cos(k\eta/\sqrt{3}) \right)\;,
\end{eqnarray}
where $k$ is the comoving wavenumber, and the primordial perturbation $\psi_k$ is obtained as
\begin{eqnarray}
\langle \psi_{\bf k}\psi_{ \tilde{\bf k}}  \rangle = \frac{2\pi^2}{k^3}\left(\frac{4}{9}\mathcal{P}_\mathcal{R}(k)\right)\delta(\bf{k}+ \tilde{\bf k})\;.
\end{eqnarray}
In the RD era, the energy density of induced GWs reads \cite{Kohri:2018}
\begin{eqnarray}\label{OGW}
&\Omega_{\rm{GW}}(\eta_c,k) = \frac{1}{12} {\displaystyle \int^\infty_0 dv \int^{|1+v|}_{|1-v|}du } \left( \frac{4v^2-(1+v^2-u^2)^2}{4uv}\right)^2\mathcal{P}_\mathcal{R}(ku)\mathcal{P}_\mathcal{R}(kv)\left( \frac{3}{4u^3v^3}\right)^2 (u^2+v^2-3)^2\nonumber\\
&\times \left\{\left[-4uv+(u^2+v^2-3) \ln\left| \frac{3-(u+v)^2}{3-(u-v)^2}\right| \right]^2  + \pi^2(u^2+v^2-3)^2\Theta(v+u-\sqrt{3})\right\}\;,
\end{eqnarray}
where $\Theta$ denotes the Heaviside theta function, and $\eta_{c}$ represents the time when $\Omega_{\rm{GW}}$ stops to
grow. The energy spectra of the induced GWs at the present time is given by \cite{Inomata:2019-a}
\begin{eqnarray}\label{OGW0}
\Omega_{\rm{GW},0}h^2 = 0.83\left( \frac{g_c}{10.75} \right)^{-1/3}\Omega_{\rm{r},0}h^2\Omega_{\rm{GW}}(\eta_c,k)\;,
\end{eqnarray}
where $\Omega_{\rm{r},0}h^2\simeq 4.2\times 10^{-5}$ is the current radiation density parameter , while $g_c\simeq106.75$ is the effective degrees of freedom in the energy density at $\eta_c$. The frequency and wavenumber can be converted to each other as
\begin{eqnarray}\label{k_to_f}
f=1.546 \times 10^{-15}\left( \frac{k}{{\rm Mpc}^{-1}}\right){\rm Hz}.
\end{eqnarray}
Now, the current energy density of the induced GWs can be obtained from Eqs. (\ref{OGW}) to (\ref{k_to_f}) and using the power spectrum estimated by the MS equation (\ref{MS_Eq}).

\begin{figure}[H]
\centering
\includegraphics[scale=0.38]{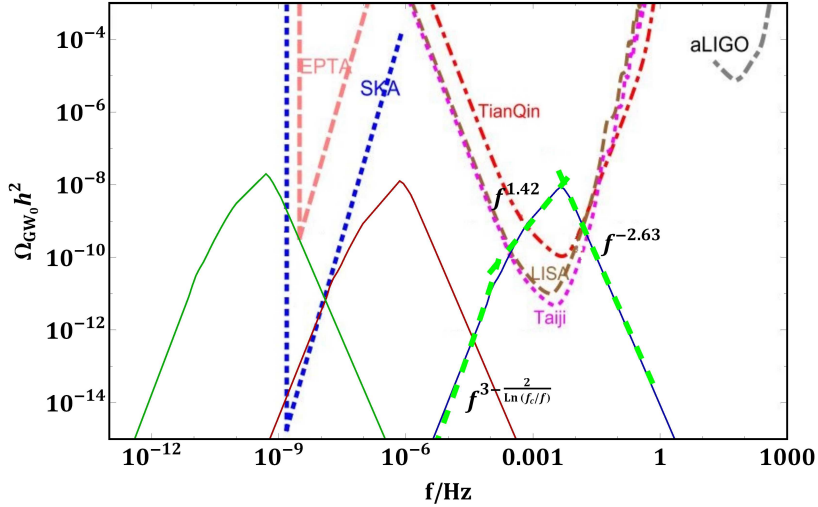}
\vspace{-0.5em}
\caption{The induced GWs energy density parameter $\Omega_{\rm GW{,0}}$ produced from the parameter sets $A$ (green line), $B$ (red line), and $C$ (blue line). The broken power-law behaviour of $\Omega_{GW}$ is shown by black dashed line. The dashed curves indicate sensitivity of GWs observatories, such as the European PTA (EPTA) \cite{EPTA-a,EPTA-b,EPTA-c,EPTA-d}, the Square Kilometer Array (SKA) \cite{ska}, the Advanced Laser Interferometer Gravitational Wave Observatory (aLIGO) \cite{ligo-a,ligo-b}, the Laser Interferometer Space Antenna (LISA) \cite{lisa,lisa-a}, Taiji \cite{taiji} and TianQin \cite{tianqin}.}
\label{fig-omega}
\end{figure}

In Fig. \ref{fig-omega}, the numerical results obtained for $\Omega_{\rm{GW},0}$ are illustrated. Figure \ref{fig-omega} shows that the induced GWs fall in the mHz band for the parameter set $C$ and can be examined by the space-based observatories, such as LISA, Taiji, and TianQin. The induced GWs generated from the parameter sets $A$ and $B$ have peaks in frequencies $f\sim10^{-10}\text{Hz}$ and $f\sim10^{-7}\text{Hz}$, respectively, and both of them can be tested by the SKA observation.

The energy density of induced GWs can be modeled by a power-law function of frequency $\Omega_{\rm GW} (f) \sim f^{n} $ \cite{Xu:2020,Fu:2020}. This parametrization can be considered as a powerful tool in probing history of the universe \cite{Kuro:2018}. For the  parameter set $C$, we obtain $\Omega_{\rm GW} \sim f^{1.42}$ for $f<f_{c}=4.42\times 10^{-3}{\rm Hz}$, and $\Omega_{\rm GW} \sim f^{-2.63}$ for $f>f_{c}$. In the infrared limit $f\ll f_{c}$, the power index is determined as $n=3-2/\ln(f_c/f)$, which is completely consistent with the analytical result obtained in \cite{Yuan:2020,shipi:2020}.

\vspace{-1.2cm}


\section{Conclusions}\label{sec7}
Here, we investigated the production of primordial black holes in a restricted $k$-inflation when the inflationary potential has exponential form. A suitable kinetic term can cause the primordial perturbations to grow up to ${\cal O}(10^{-2})$ at small scales. Additionally,  the selected function should lead to the model be consistent with Planck's observation at the pivot scale. Hence, we employ a function that includes two sections i.e $g(\phi)=g_I(\phi)\Big(1+g_{II}(\phi)\Big)$. The first part ensures that the model is compatible with Planck's observations at the CMB scale.
And the model can explain PBH formation at the small scales with the fine-tuning of the second section parameters.

We selected the three sets of parameters denoting by $A$, $B$ and $C$ for the PBH abundance $f_{\rm PBH}$.
In case $A$, $f_{\rm PBH}$ enhances to ${\cal O}(10^{-3})$, which is compatible with the upper limit of LIGO, as shown in Fig. \ref{fig-omega}. In case $B$, $f_{\rm PBH}$ falls into the allowed region of ultrashort timescale in OGLE data. For this case, we obtained the peak value of $f_{\rm PBH}$ as $\sim{\cal O}(10^{-2})$ at the mass scale $\sim{\cal O}(10^{-5})M_{\rm \odot}$. For the parameter set $C$, at the mass scale $\sim{\cal O}(10^{-13})M_{\rm \odot}$ the PBH forms by the abundance $f_{\rm PBH}\sim 0.96$. This confirms that the PBH production can justify most of DM in the universe.

In addition, we investigated the production of induced GWs for all parameter sets listed in Table\ref{tab1}. For the cases $A$ and $B$, the peaks of GWs energy density parameter are located near $10^{-10}\text{Hz}$ and $10^{-7}\text{Hz}$, respectively, which can be tested by the SKA observation. For the parameter set $C$, the peak of $\Omega_{\rm GW}$ falls into mHz band and this can be explored by the observations of LISA, Taiji, and TianQin. Our numerical calculations indicate that the GWs energy density parameter behaves like a power-law function $\Omega_{\rm GW}\sim (f/f_c)^n$. We obtained $\Omega_{\rm GW} \sim f^{2.1}$ for $f<f_{c}=3.62\times 10^{-3}{\rm Hz}$ and $\Omega_{\rm GW} \sim f^{-3.36}$ for $f>f_{c}$. In the infrared limit $f\ll f_{c}$, the power index has the logarithmic form $n=3-2/\ln(f_c/f)$, which is in good agreement with the analytical result obtained in \cite{Yuan:2020,shipi:2020}.

\subsection*{Acknowledgements}
The authors thank the referee for his/her valuable comments.



\begin{thebibliography}{0}
\expandafter\ifx\csname natexlab\endcsname\relax\def\natexlab#1{#1}\fi
\expandafter\ifx\csname bibnamefont\endcsname\relax
  \def\bibnamefont#1{#1}\fi
\expandafter\ifx\csname bibfnamefont\endcsname\relax
  \def\bibfnamefont#1{#1}\fi
\expandafter\ifx\csname citenamefont\endcsname\relax
  \def\citenamefont#1{#1}\fi
\expandafter\ifx\csname url\endcsname\relax
  \def\url#1{\texttt{#1}}\fi
\expandafter\ifx\csname urlprefix\endcsname\relax\def\urlprefix{URL }\fi
\providecommand{\bibinfo}[2]{#2}
\providecommand{\eprint}[2][]{\url{#2}}

\end{thebibliography}


\begin{thebibliography}{100}
%
\bibitem{Hawking:1971}
S. Hawking, Mon. Not. R. Astron. Soc. {\bf 152}, 75 (1971).
%
\bibitem{Carr:1974}
B. J. Carr and S.W. Hawking, Mon. Not. R. Astron. Soc. {\bf 168}, 399 (1974).
%
\bibitem{Carr:1975}
B. J. Carr, Astrophys. J. {\bf 201}, 1 (1975).
%
\bibitem{zeldovich:1967}
 Ya. B. Zel'dovich and I. D. Novikov, Sov. Astron. {\bf 10}, 602 (1967).
 %
\bibitem{Abbott:2016-a}
B. P. Abbott et al. (LIGO Scientific and Virgo Collaboration), Phys. Rev. Lett.
{\bf 116}, 061102 (2016).
%
\bibitem{Abbott:2016-b}
B. P. Abbott et al. (LIGO Scientific and Virgo Collaboration), Phys. Rev. Lett.
{\bf 116}, 241103 (2016).
%
\bibitem{Abbott:2017-a}
 B. P. Abbott et al. (LIGO Scientific and Virgo Collaboration), Phys. Rev. Lett.
{\bf 116}, 221101 (2017).
%
\bibitem{Abbott:2017-b}
B. P. Abbott et al. (LIGO Scientific and Virgo Collaboration), Astrophys. J.
{\bf 851}, L35 (2017).
%
\bibitem{Abbott:2017-c}
B. P. Abbott et al. (LIGO Scientific and Virgo Collaboration), Phys. Rev. Lett.
{\bf 119}, 141101 (2017).
%
\bibitem{Bird:2016}
 S. Bird et al., Phys. Rev. Lett. {\bf 116}, 201301 (2016).
%
\bibitem{Clesse:2017}
S. Clesse and J. Garc{\'\i}a-Bellido, Phys. Dark Universe {\bf 15}, 142 (2017).
%
\bibitem{Sasaki:2016}
M. Sasaki, T. Suyama, T. Tanaka, and S. Yokoyama, Phys. Rev. Lett. {\bf 117}, 061101 (2016).
%
\bibitem{Carr:2016}
B. Carr, F. K\"uhnel, and M. Sandstad, Phys. Rev. D {\bf 94}, 083504 (2016).
%
%
\bibitem{OGLE2}
P. Mr\'oz et al. Nature {\bf 548}, 7666 (2017).
%
\bibitem{WD}
P. W. Graham, S. Rajendran, and J. Varela, Phys. Rev. D {\bf 92}, 063007 (2015).
%
\bibitem{HSC}
H. Niikura et al., Nat. Astron. {\bf 3}, 524 (2019).
%
\bibitem{HSC-b}
 R.-G. Cai, S. Pi, S.-J. Wang, and X.-Y. Yang, JCAP {\bf 05}, 013 (2019).
%
\bibitem{sato:2019}
G. Sato-Polito, E. D. Kovetz, and M. Kamionkowski, Phys. Rev. D {\bf 100}, 063521 (2019).
%
\bibitem{akrami:2018}
Y. Akrami et al. (Planck Collaboration), A\&A {\bf 641}, A10 (2020).
%
\bibitem{Kohri:2018}
 K. Kohri and T. Terada, Phys. Rev. D {\bf 97}, 123532 (2018).
%
\bibitem{Cai:2019-a}
R.-G. Cai, S. Pi, and M. Sasaki, Phys. Rev. Lett. {\bf 122}, 201101 (2019).
%
%
\bibitem{Bartolo:2019-a}
N. Bartolo, V. De Luca, G. Franciolini, A. Lewis, M. Peloso, and A. Riotto, Phys. Rev. Lett.
{\bf 122}, 211301 (2019).
%
\bibitem{Bartolo:2019-b}
 N. Bartolo, V. De Luca, G. Franciolini, M. Peloso, D. Racco, and A. Riotto, Phys. Rev. D
{\bf 99}, 103521 (2019).
%
\bibitem{Wang:2019}
S. Wang, T. Terada, and K. Kohri, Phys. Rev. D {\bf 99}, 103531 (2019).
%
\bibitem{Cai:2019-c}
Y.-F. Cai, C. Chen, X. Tong, D.-G. Wang, and S.-F. Yan, Phys. Rev. D {\bf 100}, 043518 (2019).
%
\bibitem{Xu:2020}
W.-T. Xu, J. Liu, T.-J. Gao, and Z.-K. Guo, Phys. Rev. D {\bf 101}, 023505 (2020).
%
\bibitem{Lu:2019}
 Y. Lu, Y. Gong, Z. Yi, and F. Zhang, JCAP {\bf 12}, 031 (2019).
%
\bibitem{Hajkarim:2019}
F. Hajkarim and J. Schaffner-Bielich, Phys. Rev. D {\bf 101}, 043522 (2020).
%
\bibitem{Domenech:2020a}
G. Dom\`enech, Int. J. Mod. Phys. D {\bf 29}, 2050028 (2020).
%
\bibitem{Domenech:2020b}
G. Dom\`enech, and M. Sasaki, Phys. Rev. D {\bf 103}, 063531 (2021).
%
\bibitem{Fumagalli:2020b}
J. Fumagalli, S. Renaux-Petel, and L. T. Witkowski, arXiv:2012.02761 (2020).
%
\bibitem{Cai:2018}
Y.-F. Cai, X. Tong, D.-G. Wang, and S.-F. Yan, Phys. Rev. Lett. {\bf 121}, 081306 (2018).
%
\bibitem{Ballesteros:2019}
G. Ballesteros, J. B. Jim´enez, and M. Pieroni, JCAP {\bf 06}, 016 (2019).
%
\bibitem{Ballesteros:2020a}
 G. Ballesteros, J. Rey, M. Taoso, A. Urbano, JCAP {\bf 07}, 025 (2020).
%
\bibitem{Ballesteros:2020b}
G. Ballesteros, J. Rey, F. Rompineve, JCAP {\bf 06}, 014 (2020).
%
\bibitem{Kamenshchik:2019}
A. Y. Kamenshchik, A. Tronconi, T. Vardanyan, and G. Venturi, Phys. Lett. B {\bf 791}, 201
(2019).
%
%
\bibitem{Inomata:2018}
K. Inomata, M. Kawasaki, K. Mukaida, and T. T. Yanagida, Phys. Rev. D {\bf  97}, 043514 (2018).
%
\bibitem{Motohashi:2017}
H. Motohashi and W. Hu, Phys. Rev. D {\bf 96}, 063503 (2017).
%
\bibitem{Germani:2017}
C. Germani and T. Prokopec, Phys. Dark Univ. {\bf 18}, 6 (2017).
%
\bibitem{Di:2018}
H. Di and Y. Gong, JCAP {\bf 07}, 007 (2018).
%
%

\bibitem{Ezquiaga:2018}
J. M. Ezquiaga, J. Garc{\'\i}a-Bellido, and E. R. Morales, Phys. Lett. B {\bf 776}, 345 (2018).
%

\bibitem{Ballesteros:2018}
G. Ballesteros and M. Taoso, Phys. Rev. D {\bf 97}, 023501 (2018).
%
\bibitem{Dalianis:2019}
I. Dalianis, A. Kehagias, and G. Tringas, JCAP {\bf 01}, 037 (2019).
%
\bibitem{Dalianis:2020}
 I. Dalianis, S. Karydas, and E. Papantonopoulos, JCAP {\bf 06}, 040 (2020).
%
\bibitem{chen:2019}
C. Chen and Y.-F. Cai, JCAP {\bf 10}, 068 (2019).
%
\bibitem{Ozsoy:2018}
O. \"Ozsoy, S. Parameswaran, G. Tasinato, and I. Zavala, JCAP {\bf 07}, 005 (2018).
%
\bibitem{Tada:2019}
Y. Tada and S. Yokoyama, Phys. Rev. D {\bf 100}, 023537 (2019).
%
\bibitem{Liu:2020}
J. Liu, Z.-K. Guo, and R.-G. Cai, Phys. Rev. D {\bf 101}, 023513 (2020).
%
\bibitem{Bellido:2017}
J. Garc{\'\i}a-Bellido and E. R. Morales, Phys. Dark Univ. {\bf 18}, 47 (2017).
%
%
%
\bibitem{mishra:2020}
S. S. Mishra and V. Sahni, JCAP {\bf 04},  007 (2020).
%
\bibitem{Fumagalli:2020a}
J. Fumagalli, S. Renaux-Petel, J. W. Ronayne, and L. T. Witkowski, arXiv:2004.08369.
%
\bibitem{Atal:2019}
V. Atal, J. Garriga, and A. Marcos-Caballero, JCAP {\bf 09} 073 (2019).
%
\bibitem{Khlopov:2010}
M. Yu. Khlopov, Res. Astron. Astrophys. {\bf 10}, 495 (2010).
%
\bibitem{Belotsky1:2014}
K. M. Belotsky et al., Mod. Phys. Lett. A, {\bf  29}, 1440005 (2014).
%
\bibitem{Belotsky:2019}
K. M. Belotsky et al., Eur. Phys. J. C, {\bf  79}, 246 (2019).
%
\bibitem{shiPi:2018}
S. Pi, Y.-l. Zhang, Q.-G. Huang and M. Sasaki, JCAP {\bf 05}, 042 (2018).
%
\bibitem{Braglia:2020}
M. Braglia, D. K. Hazra, F. Finelli, G. F. Smoot, L. Sriramkumar and A. A. Starobinsky, JCAP {\bf 08}, 001 (2020).
%
\bibitem{Braglia2:2020}
M. Braglia, X. Chen, and D. K. Hazra, JCAP {\bf 03}, 005 (2021).
%
\bibitem{mahbub:2020}
R. Mahbub, Phys. Rev. D {\bf 101}, 023533 (2020).
%

\bibitem{Tsujikawa:2011}
 A. De Felice, S. Tsujikawa, Phys. Rev. D {\bf 84}, 083504 (2011).
%
\bibitem{Tsujikawa:2011-b}
A. De Felice, S. Tsujikawa, J. Elliston, R. Tavakol, JCAP {\bf 08}, 021 (2011).
%
\bibitem{teimoori:2018}
 Z. Teimoori and K. Karami, Astrophys. J. {\bf 864}, 41 (2018).
%
\bibitem{kobayashi:2010}
 T. Kobayashi, M. Yamaguchi, J. Yokoyama, Phys. Rev. Lett. {\bf 105}, 231302 (2010).
%
\bibitem{Ohashi:2012}
J. Ohashi and S. Tsujikawa, JCAP {\bf 10}, 035 (2012).
%
\bibitem{Tumurtushaa:2019}
G. Tumurtushaa,  Eur. Phys. J. C {\bf 79}, 920 (2019).
%
\bibitem{Burrage:2010}
C. Burrage, C. d. Rham, D. Seery, and A. J. Tolley, JCAP {\bf 01}, 014 (2011).
\bibitem{Horndeski:1974}
G. W. Horndeski, Int. J. Theor. Phys. {\bf 10}, 363 (1974).
%
\bibitem{Hirano:2016}
S. Hirano, T. Kobayashi and S. Yokoyama, Phys. Rev. D {\bf 94}, 103515 (2016).
%
\bibitem{Armendariz:1999}
C. Armendariz-Picon, T. Damour and V. F. Mukhanov, Phys. Lett. B {\bf 458}, 209 (1999).
%
\bibitem{Garriga:1999}
J. Garriga and V. F. Mukhanov, Phys. Lett. B {\bf 458}, 219 (1999).
%
\bibitem{lin:2020}
J. Lin, Q. Gao, Y. Gong, Y. Lu, C. Zhang, and F. Zhang, Phys. Rev. D {\bf 101}, 103515 (2020).
%
\bibitem{Barenboim:2007}
G. Barenboim and W. H. Kinney, JCAP {\bf 03}, 014 (2007).
%
\bibitem{karami:2017}
 Z. Teimoori, K. Karami, Nucl. Phys. B {\bf 21}, 25 (2017).
%
\bibitem{zhang:2021}
F. Zhang, Y. Gong, J. Lin, Y. Lu and Z. Yi, JCAP {\bf 04}, 045 (2021).
%
\bibitem{yi:2021}
Z. Yi, Q. Gao, Y. Gong and Z.-H. Zhu, Phys. Rev. D {\bf 103}, 063534 (2021).
%
\bibitem{Inomata:2019-a}
 K. Inomata and T. Nakama, Phys. Rev. D {\bf 99}, 043511 (2019).
%
\bibitem{Inomata:2016}
K. Inomata, M. Kawasaki, and Y. Tada, Phys. Rev. D {\bf 94}, 043527 (2016).
%
\bibitem{Fixsen:1996}
 D. J. Fixsen, E. S. Cheng, J. M. Gales, J. C. Mather, R. A. Shafer, and E. L. Wright, Astrophys. J. {\bf 473}, 576 (1996).
%
%
\bibitem{Fu:2019} C. Fu, P. Wu, and H. Yu, Phys. Rev. D {\bf 100}, 063532 (2019).

\bibitem{Sasaki:2018}
M. Sasaki, T. Suyama, T. Tanaka and S. Yokoyama, Class. Quantum Grav. {\bf 35}, 063001 (2018).
%
\bibitem{Salopek:1998}
D. S. Salopek, J. R. Bond and J. M. Bardeen, Phys. Rev. D {\bf 40}, 1753 (1989).
%
\bibitem{Mukhanov:1991}
 V. F. Mukhanov and M. I. Zelnikov, Phys. Lett. B {\bf 263}, 169 (1991).
%
\bibitem{Starobinsky:1992}
A. A. Starobinsky, JETP Lett. {\bf 55}, 489 (1992)
%
\bibitem{Bellido:1996}
J. Garc{\'\i}a-Bellido, A. Linde and D. Wands, Phys. Rev. D {\bf 54}, 6040 (1996).
%
\bibitem{Randall:1996}
 L. Randall, N. Sikjacic and A. Guth, Phys. Lett. B {\bf 472}, 377 (1996).
%
\bibitem{Ivanov:1994}
P. Ivanov, P. Naselsky  and I. Novikov, Phys. Rev. D {\bf 50}, 7173 (1994)
%
\bibitem{Ivanov:1998}
P. Ivanov, Phys. Rev. D {\bf 57}, 7145 (1998)
%
\bibitem{Inomata:2017}
K. Inomata, M. Kawasaki, K. Mukaida, Y. Tada and T. T. Yanagida, Phys. Rev. D {\bf 96}, no. 4, 043504 (2017)
%
\bibitem{young:2014}
S. Young, C. T. Byrnes and M. Sasaki, JCAP {\bf 07}, 045 (2014).
%
\bibitem{harada:2013}
 T Harada, C.-M. Yoo and K. Kohri, Phys. Rev. D {\bf 88}, 084051 (2013).
%
\bibitem{Musco:2013}
I. Musco and J. C. Miller, Class. Quant. Grav. {\bf 30}, 145009 (2013).
%
\bibitem{Germani:2019}
C. Germani and I. Musco, Phys. Rev. Lett. {\bf 122}, no. 14, 141302 (2019)
%
\bibitem{Shibata:1999}
M. Shibata and M. Sasaki, Phys. Rev. D {\bf 60} (1999) 084002.
%
\bibitem{Polnarev:2007}
A. G. Polnarev and I. Musco, Class. Quant. Grav. {\bf 24}, 1405 (2007).
%
\bibitem{Musco:2009}
I. Musco, J. C. Miller and A. G. Polnarev, Class. Quant. Grav. {\bf 26}, 235001 (2009).
%
\bibitem{Ali:2017}
Y. Ali-Ha{\"i}moud, E. D. Kovetz, and M. Kamionkowski, Phys. Rev. D \textbf{96}, 123523 (2017).
%
\bibitem{EGG}
B. J. Carr, K. Kohri, Y. Sendouda, and J. Yokoyama, Phys. Rev. D {\bf 81}, 104019 (2010)
%
\bibitem{femto}
A. Barnacka, J. F. Glicenstein, and R. Moderski, Phys. Rev. D {\bf 86}, 043001 (2012).
%


\bibitem{kepler}
 K. Griest, A. M. Cieplak, and M. J. Lehner, Phys. Rev. Lett. {\bf 111}, 181302 (2013).
%
\bibitem{EROS}
 P. Tisserand et al. (EROS-2 Collaboration), Astron. Astrophys. {\bf 469}, 387 (2007).
%
\bibitem{CMB-a}
Y. Ali-Ha{\"i}moud and M. Kamionkowski, Phys. Rev. D \textbf{95}, 043534 (2017).
%
\bibitem{CMB-b}
V. Poulin, P. D. Serpico, F. Calore, S. Clesse, and K. Kohri, Phys. Rev. D {\bf 96}, 083524 (2017).
%
\bibitem{EPTA-a}
R. D. Ferdman et al., Class. Quant. Grav. {\bf 27}, 084014 (2010).
%
%
\bibitem{EPTA-b}
G. Hobbs et al., Class. Quant. Grav. {\bf 27}, 084013 (2010).
%
\bibitem{EPTA-c}
M. A. McLaughlin, Class. Quant. Grav. {\bf 30}, 224008 (2013).
%
\bibitem{EPTA-d}
G. Hobbs, Class. Quant. Grav. {\bf 30}, 224007 (2013).
%
\bibitem{lisa}
P. Amaro-Seoane et al. (LISA Collaboration), arXiv:1702.00786.

\bibitem{Ananda:2007}
K. N. Ananda, C. Clarkson, and D. Wands, Phys. Rev. D \textbf{75}, 123518 (2007).
%
\bibitem{Baumann:2007}
D. Baumann, P. J. Steinhardt, K. Takahashi and K. Ichiki, Phys. Rev. D \textbf{76}, 084019 (2007).
%



\bibitem{Fu:2020}
C. Fu, P. Wu, and H. Yu, Phys. Rev. D {\bf 101}, 023529 (2020).






%
\bibitem{Kuro:2018}
S. Kuroyanagi, T. Chiba, and T. Takahashi, JCAP {\bf 11}, 038 (2018).
%
\bibitem{Yuan:2020}
C. Yuan, Z.-C. Chen, and Q.-G. Huang, Phys. Rev. D {\bf 101}, 043019 (2020).
%
\bibitem{shipi:2020}
R.-G. Cai, S. Pi, and M. Sasaki, Phys. Rev. D {\bf 102}, 083528 (2020).
%
\bibitem{ska}
C. J. Moore, R. H. Cole, and C. P. L. Berry, Class. Quant. Grav. {\bf 32}, 015014 (2015).
%
\bibitem{ligo-a}
G. M. Harry (LIGO Scientific Collaboration), Class. Quant. Grav. {\bf 27}, 084006 (2010).
%
\bibitem{ligo-b}
J. Aasi et al. (LIGO Scientific Collaboration), Class. Quant. Grav. {\bf 3}2, 074001 (2015).
%
\bibitem{lisa-a}
K. Danzmann, Class. Quant. Grav. {\bf 14}, 1399 (1997).
%
\bibitem{taiji}
W.-R. Hu and Y.-L. Wu, Natl. Sci. Rev. {\bf 4}, 685 (2017).
%
\bibitem{tianqin}
J. Luo et al. (TianQin Collaboration), Class. Quant. Grav. {\bf 33}, 035010 (2016).
%
\end{thebibliography}
\end{document}